\documentclass[galaxies,review,accept,pdftex,oneauthor]{Definitions/mdpi}

\newcommand{\WHz}{$\,$W$\,$Hz$^{-1}$}

\newcommand{\upmuJybeam}{$\upmu$Jy beam$^{-1}$}

\newcommand{\alwmath}[1]{\ifmmode#1\else$#1$\fi}
\newcommand{\ltsima} {$\; \buildrel < \over \sim \;$}
\newcommand{\gtsima} {$\; \buildrel > \over \sim \;$}
\newcommand{\lta} {\lower.5ex\hbox{\ltsima}}
\newcommand{\gta} {\lower.5ex\hbox{\gtsima}}

\usepackage{longtable}

\firstpage{1} 
\pubvolume{12}
\issuenum{2}
\articlenumber{11}
\pubyear{2024}
\copyrightyear{2024}
\externaleditor{Academic Editor: Fulai Guo}
\datereceived{2 February 2024} 
\daterevised{4 March 2024} 
\dateaccepted{5 March 2024} 
\datepublished{13 March 2024} 
\hreflink{https://doi.org/10.3390/\linebreak galaxies12020011} 

\Title{{What Have We Learned about} the Life Cycle of Radio {Galaxies} from New Radio Surveys} 

\TitleCitation{What Have We Learned about the Life Cycle of Radio {Galaxies} from New Radio Surveys}

\Author{{Raffaella Morganti} 
 $^{1,2}$\orcidA{}}

\AuthorCitation{Morganti, R.}

\address{%
$^{1}$ \quad ASTRON{-} 
 The Netherlands Institute for Radio Astronomy, Oude Hoogeveensedijk 4,\linebreak  7991 PD Dwingeloo, {The Netherlands; morganti@astron.nl} 
\\
$^{2}$ \quad Kapteyn Astronomical Institute, University of Groningen, Landleven 12,\linebreak 9747 AD Groningen, The Netherlands}

\abstract{The recurrent activity of radio AGN, with phases of activity alternating with periods of quiescence, has been known since the early studies of these objects. The full relevance of this cycle is emphasised by the requirement, from  the AGN feedback scenario, of a recurrent impact of the energy released by the SMBH during the lifetime of the host galaxy: only in this way can AGN feedback influence galaxy evolution.  
Radio AGN in different evolutionary phases can be identified by their properties, like morphology and spectral indices. Dying/remnant and restarted sources have been the most elusive to select and characterise, but they  are crucial to quantify the full life cycle.
Thanks to the availability of new, large radio surveys (particularly at low frequencies), it is finally possible to make a more complete census of these rare sources and start building larger samples. 
This paper gives an overview of the recent work conducted using a variety of radio telescopes and surveys,  highlighting some of the new results characterising the properties of dying/remnant and restarted radio sources and what has been learned about the life cycle of radio AGN. The comparison with the predictions from numerical simulations is also discussed. 
The results so far show that remnant and restarted radio AGN have a variety of properties which make these objects more complex than previously thought.
}


\begin{document}

\section{Introduction}

A fascinating aspect of radio active galactic nuclei  (AGN) is that their evolutionary stage can be identified and timed. This is something quite unique for AGN and extremely valuable not only for understanding the processes that trigger or halt the activity but~also for assessing the relevance of these phases in the context of AGN feedback. Recurring nuclear activity during the life of a galaxy is a requirement of the cosmological simulations in order to limit, by~regulating accretion and the amount of gas present, the~growth of super-massive black holes (SMBH) and the star formation in massive galaxies \cite{Silk98,Fabian12,Gaspari17}{.} 

\begin{figure}[H]
\hspace{-16pt}\includegraphics[width=13.5 cm]{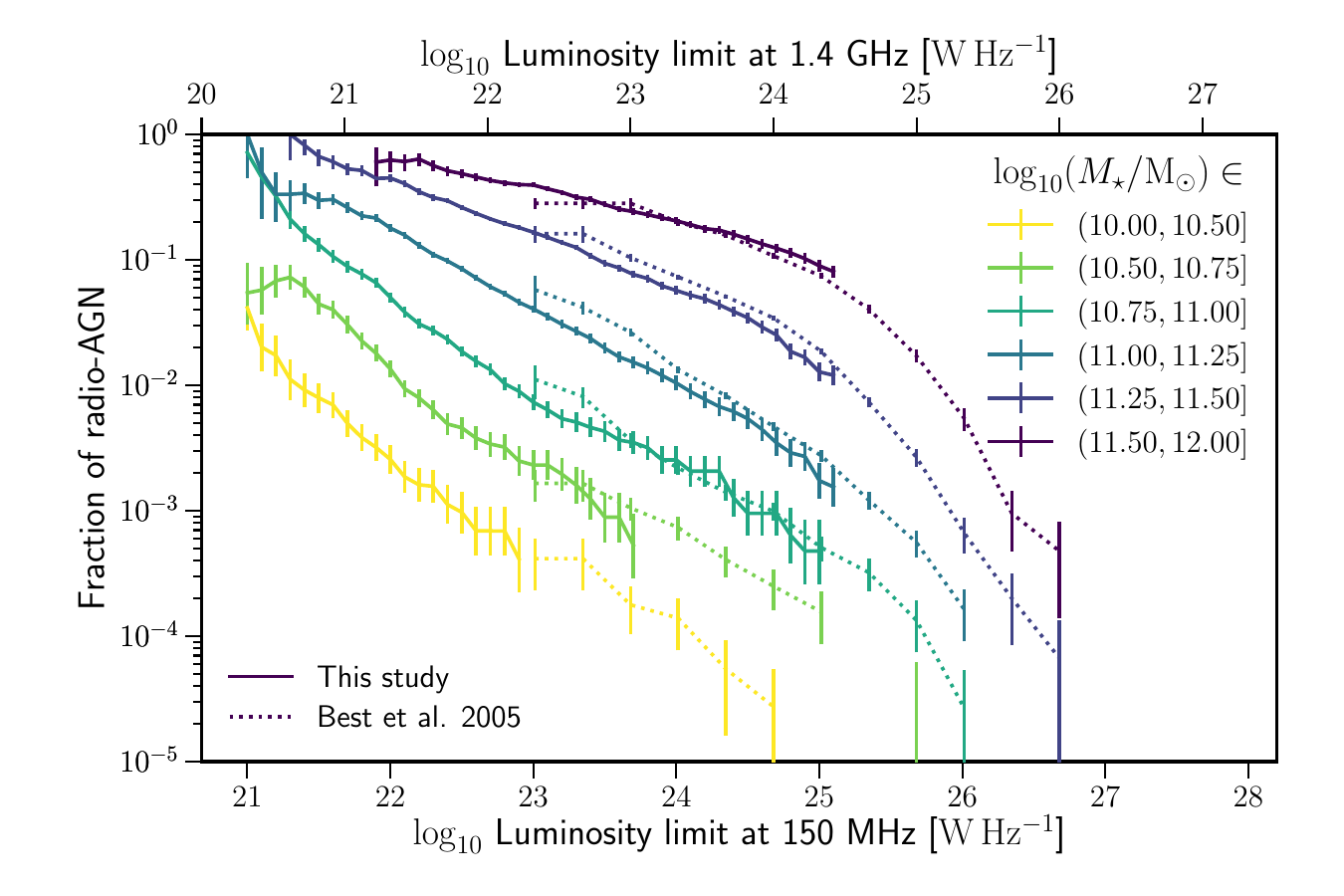}
\caption{Fraction of galaxies hosting radio AGN brighter than a given radio luminosity, separated by their stellar mass (shown in different colours). Figure taken from~\cite{Sabater19}.  {{Radio emission from AGN was separated from starformation using a number of criteria, see  paper for details.}} The solid lines represent the results obtained at 150~MHz (\cite{Sabater19}), and dotted lines show the results from~\cite{Best05} at 1.4~GHz (converted assuming a spectral index of 0.7). 
Reproduced with permission from Astronomy \& Astrophysics, \copyright ESO.
\label{fig:fig1}}
\end{figure}

Different approaches have been taken to explore the life cycle of radio AGN. For~example, some studies~\cite{Best05,Sabater19} have looked at the fraction of galaxies that host a radio AGN above a given luminosity limit. To~the first order (and under the assumption that every massive galaxy becomes a radio source during its life), this fraction can be used as a proxy for the fraction of time a radio AGN is “on”  (i.e., active). Figure~\ref{fig:fig1} (taken from~\cite{Sabater19}) shows that the fraction of low-power radio sources is higher, suggesting that the most massive galaxies host always a radio AGN. This indicates a different and likely shorter duty cycle compared to the one of less massive galaxies \cite{Best05,Sabater19}. 
{{Interestingly,} 
 the~study of~\cite{Capetti22}, while confirming these results, also finds radio sources hosted by massive galaxies which are in the remnant stage: this suggests that the actual evolution of these objects can be more complex and highlights the relevance of detailed studies.}
The presence of multiple phases of activity is also seen imprinted in the X-ray cavities observed in cool-core clusters, groups and in the halo of massive early-type galaxies (see, e.g.,~\cite{Randall11,Vantyghem14,Biava21}). The~X-ray cavities can show the impact of multiple episodes of radio emission, even when this emission has already faded away (i.e., ghost cavities,~\cite{Fabian12}). However, these studies require extremely deep X-ray observations available only for a limited group of~objects.
\vspace{-8pt}

These results show that the full characterisation of the life cycle of radio AGN requires to derive the physical properties of objects in all phases of evolution, i.e.,~young, evolved, dying (which will be indicated as {{remnant}
} throughout this paper) and restarted sources. 
This has been historically performed using the spectral properties of the sources and, in~some cases, using their morphology. A~common way of estimating the ages of radio AGN 
is based on the theory of synchrotron (see~\cite{Harwood13} and references therein), the~mechanism at the origin of the radio emission in these objects. The~properties of the radio spectra can be used to select, for example,~young sources, indicated by their peaked spectra due to synchrotron self-absorption or free--free absorption  (e.g.,~\cite{Fanti90,ODea98,ODea21}), or~remnant radio sources, indicated by an ultra-steep spectrum of aged electrons due to energy losses (radiative and adiabatic expansion, see~\cite{Komissarov94,Murgia11,Harwood13}). The~morphological properties have been used to identify some types of restarted radio sources (i.e., double--double radio galaxies, DDRG~\cite{Schoenmakers00}).

{{However, radio sources in the remnant phase (when the activity from the central SMBH stops or dims) and in the restarted phase (when the SMBH activity starts again) have been the most elusive to identify and it has been difficult until recently to} 
} build large samples and obtain a good census of their properties. This is a major limitation because without understanding more about these phases, we cannot time the full life cycle of radio AGN. 
Because the {{observations are so time consuming}}, the~studies of these objects have been, until~recently, mostly limited to single cases, often serendipitously discovered, or~to small samples (see~\cite{Cordey87,Murgia11,Konar13,Brienza16,Shulevski17,Brienza20,Maccagni20,Kukreti22,Candini23} to mention some examples).  
However, we have now entered a new era with new possibilities to select remnant and restarted radio sources over a larger area of the sky and using multiple selection criteria. This has provided a boost in the field with exciting new results as~described in this short review. 
For interested readers, the~overview in this paper can be complemented by a number of recent reviews on radio galaxies, including aspects of  their life cycle, see~\cite{Morganti17,Hardcastle20,Morganti21a,ODea21,Mahatma23}. 

\section{Using Multiple Criteria to Select Samples of Remnants and Restarted Radio~Galaxies}
\label{sec:sec2}

In the past years,  new (or upgraded) radio telescopes have become available, and new surveys have been produced. They have made it possible to start building samples, larger than ever before, of~rare objects, like remnant and restarted radio sources. In~particular, we now have~the following:  
\begin{itemize}
\item Deep low-frequency ($\lesssim$200 MHz) surveys (i.e., the LOFAR Two-metre Sky Survey LoTSS,~\cite{Shimwell20}; the Galactic and Extra-Galactic All-Sky MWA Survey (GLEAM), \cite{Hurley22}, and TIFR-GMRT Sky Survey (TGSS)  \cite{Intema17}), tracing low surface-brightness radio structures. This emission is particularly sensitive to the location of old (remnant) electrons.
\item  Surveys at higher frequencies (e.g.,  at 1.4~GHz using the APERture Tile In Focus (Apertif) on the WSRT telescope \citep{Adams22} and the Australian Square Kilometre Array Pathfinder (ASKAP) \cite{McConnell20}) and surveys at high spatial resolution (e.g.,  at 3~GHz, the Very Large Array Sky Survey (VLASS), 
{\cite{Lacy20,Gordon21}}). The~latter is particularly useful for characterising the spectral properties of the central regions (e.g.,  cores). 
\item Spectral {indices}
 \endnote{In this paper, the~spectral index $\alpha$ is defined through $S\propto \nu^{-\alpha}$.}  extended to low frequencies can be derived over large areas of {{the}} sky. Furthermore, by~combining some of the surveys, spectral index images can \mbox{be obtained}. 
\item  Availability of high-quality ancillary data (e.g.,  optical ID and redshift) for larger areas of {{the}} sky (e.g.,  {\cite{Duncan21,Hardcastle23,DESI23}}) as well as improved methods for deriving photometric redshifts for a large number of galaxies~\cite{Duncan21}. 
\end{itemize}

The selection of remnant and restarted radio sources was performed in the past, using, as~mentioned earlier, a~limited set of spectral and/or morphological features. In~practice, this meant limiting the selection to groups of objects likely reflecting particular conditions or evolutionary path, e.g.,~DDRG for restarted and ultra-steep spectrum (USS, i.e.,~spectral index steeper than 1.2) sources for remnants.
However, single object studies have already shown us that the situation is more complex and that radio AGN can follow different evolutionary paths. Cases like the remnant nicknamed {\sl {blob 1} }, where the USS component is seen only at high frequencies \cite{Brienza16}{,} or~restarted sources only identified thanks to the distribution of the spectral indices \cite{Roettiger94,Brienza20}{,} show the variety of situations that can be~encountered. 

{{Thanks to the wealth of (radio) data available, this variety of properties can now be captured by using multiple criteria for selecting remnants and restarted candidates. This broader approach results in a more complete census of these objects.}}
Thus, the~studies that are presented below use, as much as possible, a combination of multiple criteria derived from the studies of single objects. The~{{ones}} used to select remnant radio sources are (i)~morphology (amorphous/low surface brightness, see also~\cite{Saripalli12}); (ii) absence of core (or low core prominence, defined as {CP} 
 = S$_{core}$/S$_{total}$, \cite{Giovannini88}); (iii) ultra-steep spectrum emission (USS, e.g.,~\cite{Parma07,Murgia11}); and (iv) spectral break (or high spectral curvature). {{In a similar way}}, the~parameters that can be used to select restarted radio sources are
(i) morphology \mbox{(e.g.,  double--double);} (ii) high core prominence combined with low surface brightness lobes, suggesting the presence of an active, newly started radio source surrounded by a remnant emission from the previous active phase;
(iii) peaked or steep spectrum core, indicating synchrotron self-absorbed (or free--free absorption) emission typical of young radio \mbox{sources \cite{ODea21}}{;}
and (iv) resolved spectral index showing two well-distinct populations (one of which is USS tracing a past phase of activity~\cite{Roettiger94,Brienza20}).

\unskip

\section{Selecting Samples of Remnant and Restarted Radio~Sources}

A few studies aimed at selecting and characterising samples of remnant and restarted radio sources have been recently published. These projects are focused on famous, well-studied regions (i.e., relatively ``small'' areas between 5 and 20 sq deg) because~of the wealth of ancillary data available. They represent a first step forward in the field, and below is a summary of some of their new results.
{{For reference with respect to Figure}~\ref{fig:fig1}, {the~sources included in the samples described below have low-frequency radio luminosity between a few} $\times 10^{24}$ {and a few} $\times 10^{27}$  \WHz}.

\subsection{Remnants Radio~Sources}

As mentioned above, the~selection of remnant radio sources used to be mostly limited to USS, with~a minority of studies looking for the lack of cores.  The~use of the morphology and, in~particular, the~presence of low surface brightness structures, has been pioneered by the work of~\cite{Saripalli12}, which, however, was limited to observations at 1.4~GHz. This selection can be more efficiently performed at low frequencies (i.e., frequencies well below 1~GHz), where the low-surface brightness structures can be more easily traced. 
LOFAR, providing a combination of relatively high spatial resolution and high sensitivity to low surface brightness structure, has given a unique contribution. 
An example of this is the study presented in~\cite{Brienza17}, which has selected remnant radio sources in the Lockman Hole (LH) region imaged by LOFAR at 150~MHz \cite{Tasse21}.  
For the selection, they used the morphology and the properties of the spectral indices. This study (and the follow up at 6~GHz with the VLA,~\cite{Jurlin21}) identified 13 remnant sources in this area, corresponding to about 7\% of the original sample made of sources larger than 60 arcsec. 
Interestingly, only 4.1\% of the remnant sources are USS, confirming the relatively low fraction of USS even after extending the search to low frequencies. Thus, this work has shown that the amorphous, low surface brightness morphology can provide the larger contribution of selected remnants and, crucial for future searches, not all remnants sources are USS at low frequencies.
While the selection was not based on the lack of cores, the~candidate remnants all show a low CP. Surprisingly, Ref.~\cite{Jurlin21} finds cases of remnants with USS total emission where the core is detected by the deeper 6~GHz observations (see Section \ref{sec:Cores}). 

The finding of a variety of spectral indices of the remnant structures is interesting. This result, confirmed by other studies (see below), suggests that remnant sources must be going through an evolution. This was further investigated and confirmed by some simple modelling; see Section \ref{sec:Intepretation}. A~follow-up study performed in the same region has derived resolved spectral indices and showed that in some of the remnant sources, the entire emission, which can extend a few hundred kpc, is USS,~\cite{Morganti21a,Morganti21b}. These are likely the older remnants still visible by our radio telescopes. 
Finally, unlike previous studies, only a minority of the selected remnant sources are in a cluster~environment.


The study of~\cite{Mahatma18} focused instead on the Herschel-ATLAS region and used the core-undetected method to identify remnant radio sources. This required follow-up VLA 6~GHz high-resolution observations of the objects pre-selected in the LOFAR image by~\cite{Hardcastle16}. They derived an upper limit on the fraction of remnants of about 9\%, consistent with~\cite{Brienza17}.  They also found that the remnant structures have a wide range of spectral indices \mbox{($0.5<\alpha_{1400}^{150}< 1.5$)}. This confirms that the lobes of some remnants may be characterised by spectra indices at low frequencies similar to the one of active sources, suggesting that the steepening is occurring only at higher frequencies and has not yet reached and affected the low~frequencies. 

\begin{figure}[H]
\begin{adjustwidth}{-\extralength}{0cm}
\centering
\hspace{-6pt}\includegraphics[width=16.5cm]{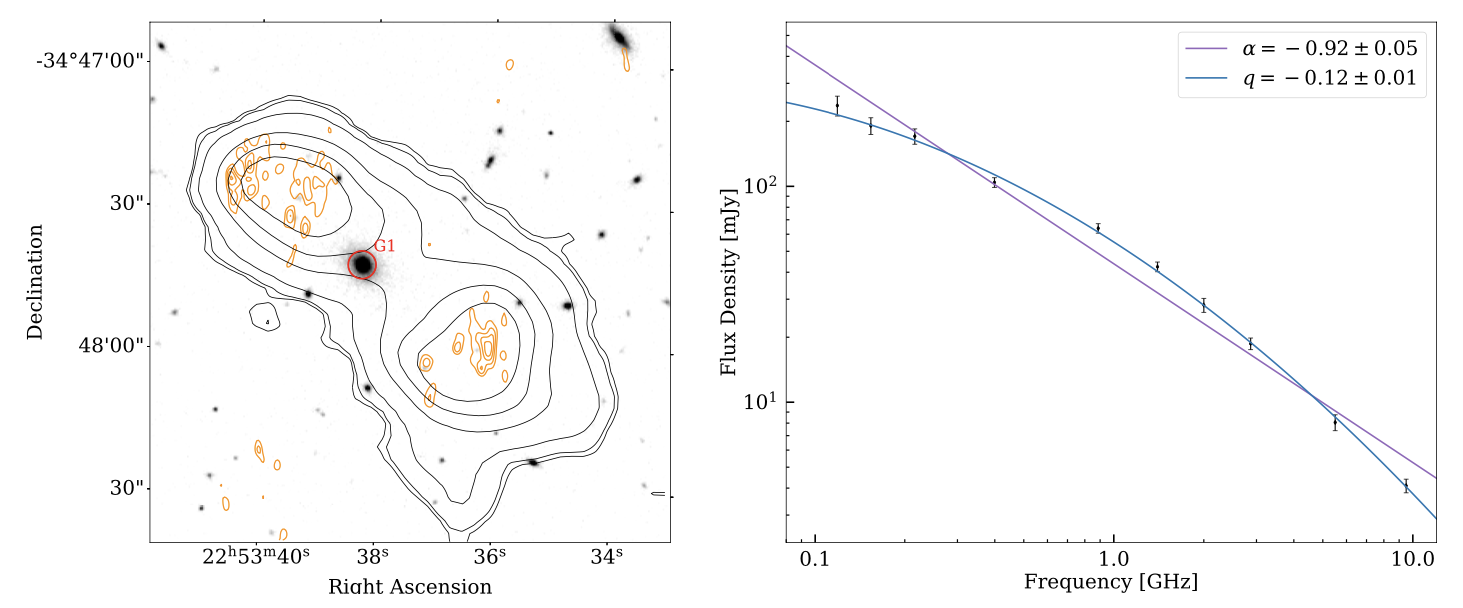}
\end{adjustwidth}
\caption{\label{fig:Quici} 
\textls[-10]{{The} continuum image (left) and integrated spectrum (right) of MIDAS J225543$-$344047 \mbox{from~\cite{Quici21}.} 
Background image is a VIKING Ks band cutout set on a linear stretch. Three sets of contours are overlaid, representing the radio emission as seen by ASKAP EMU$-$ES (black contours, levels: [3, 4, 5, 7, 15, 30, 100] $\times~\sigma$), ATCA GLASS 5.5 GHz (orange contours, levels: [3, 5, 10, 20] $\times~\sigma$). Red marker indicates the positions of the host {{galaxy}}. Right: The radio continuum spectrum between \mbox{119 MHz} and \mbox{9 GHz}. A~simple power-law  and curved power-law model are fit to the spectrum, indicated by the purple and blue models, respectively. See~\cite{Quici21} for details. Copyright \copyright\ 2021, Cambridge \mbox{University Press.}}}

\end{figure}

Quici~et~al. (2021) looked at the GAMA23 region using MWA, ASKAP and ATCA to identify sources which lack radio cores. These selection criteria resulted in 10 remnant candidates, corresponding to a fraction of remnants between 4\% and 10\%. One of their best cases in presented in Figure~\ref{fig:Quici}. Of~the selected remnant sources, only {{three}} appear to be USS.  This study also confirmed that cores can appear when deeper observations are available.   
Interestingly, they also found (like~\cite{Jurlin21}) that objects with core can have aged spectra (USS). As~in the case of~\cite{Mahatma18}, hot spots are seen in some objects: based on their spectra properties, the~authors derived that these structures can persist 5--10 Myr after the emission from the core has switched~off. 

Finally, Ref.~\cite{Dutta23} searched for remnant sources in the XMM-LSS field (using images from GMRT, LOFAR and VLA for a subregion), covering an area between \mbox{5 and 12 sq deg}.
Using both the undetected radio cores method as well as the high spectral curvature and ultra-steep spectrum, they identified 21 remnant candidates (resulting in $\sim$8\% of remnant sources). Like in~\cite{Jurlin21}, remnant sources were found to reside mostly in non-cluster environments. As~noted also in other studies,  the~selected remnant sources exhibit a variety of properties in terms of morphology, spectral index (in the range of 0.75 to 1.71), and~linear radio size (ranging from 242 kpc--1.3 Mpc).  Similar to other studies, only $\sim$3.9\% of the sources can be identified as USS~sources.


\subsection{The Presence of Cores in Candidates~Remnant} 
\label{sec:Cores}

\begin{figure}[H]
\begin{adjustwidth}{-\extralength}{0cm}
\centering
\includegraphics[width=18cm]{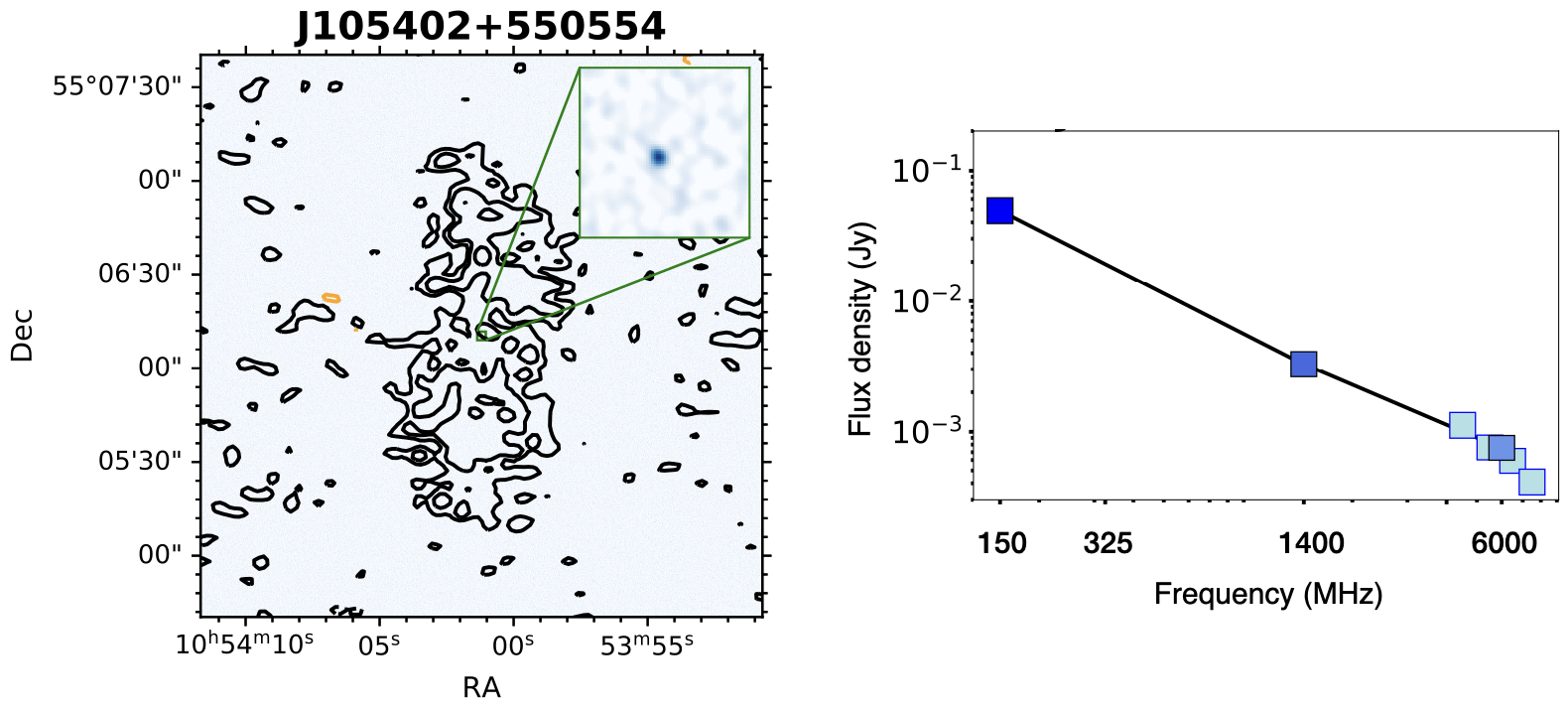}
\end{adjustwidth}
\caption{\label{fig:coreNika} Radio image (left) of one of the remnants selected in~\cite{Brienza17} and followed up at higher frequencies and resolution by~\cite{Jurlin21}. The~green square represents the position of the optical host galaxy and show the detection of a corresponding nuclear emission at 6 cm using the VLA (\mbox{S$_{\rm core, 6 GHz} = 0.142$ mJy}). The~contours of the LOFAR 6 arcsec image are shown in black. 
The right panel shows the radio spectra of the total emission, derived by using fluxes from LoTSS, GMRT, NVSS and VLA (see~\cite{Jurlin21} for details). Despite the presence of the core, the~extended emission is USS indicative of a remnant emission. 
Taken from~\cite{Jurlin21}, see text for details. Reproduced with permission from Astronomy \& Astrophysics, \copyright ESO.}

\end{figure}

One of the unexpected properties of the remnant sources is the finding, when deep enough observations are available, of~faint core emission in (some of the) objects characterised by USS extended emission. This was already seen in the detailed studies of the \mbox{{\sl {blob 1}}} and NGC~507 (see~\cite{Brienza16,Brienza22}, respectively) but it is now found to be more common. 
A nice example obtained from the sample of remnants studied by~\cite{Brienza17,Jurlin21} is the source shown in Figure~\ref{fig:coreNika}. In~this source, a core is detected at 6~GHz using A-array observations with the VLA  while no large-scale jets are observed, and the extended emission has a steep spectral index, $\alpha_{1400}^{150}\sim1.2$. 

Explaining this finding provides an interesting puzzle, and two possibilities have been put forward. Either---at least in some cases---the cores never switch off completely but~the emission only dims, reaching the very low CP characteristics of remnant sources, or, alternatively, these faint cores represent the restarting of the nuclear activity and a new phase for the radio source. If~the faint cores represent a restarted activity in a remnant structure, this  would suggest that in some objects, the ``off'' time is relatively short; see Section \ref{sec:Intepretation}. This is also suggested by the properties of the spectral index of some sources (see also Section \ref{sec:restarted}).

Finally, it is interesting to note that when faint cores appear from deep, high-resolution radio observations, they help in confirming the identifications (made with other methods) of the host galaxy of the remnant radio structure. Because~of their amorphous structure, it is usually difficult to find the optical identification of remnant radio sources and methods based on the barycenter of the emission used (see ~Ref.~\cite{Jurlin21} for some examples and discussion). Thus, the~confirmation {{a posteriori}
} of the optical identification using the faint cores gives more confidence in this~method.

\subsection{Candidates Restarted Radio~Sources}
\label{sec:restarted}

The selection of candidate restarted radio sources remains challenging and detailed studies are needed to test the criteria listed in Section \ref{sec:sec2}. Despite these challenges, a~number of interesting results have been recently~reported.

\begin{figure}[H]
\begin{adjustwidth}{-\extralength}{0cm}
\centering
\includegraphics[width=18cm]{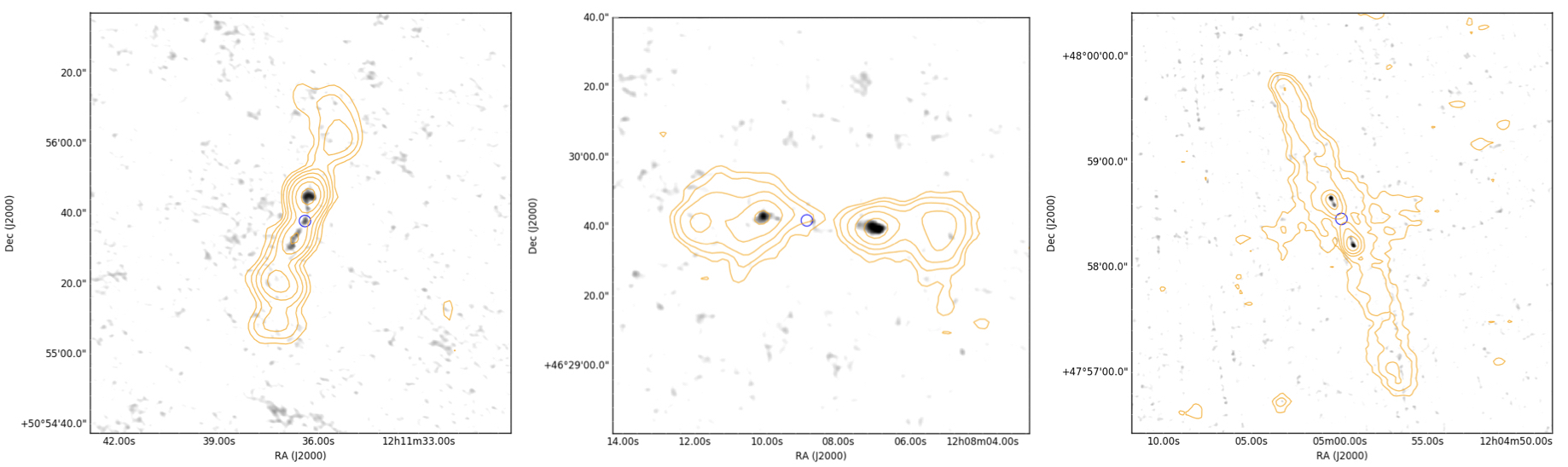}
\end{adjustwidth}
\caption{\label{fig:DoubleDouble} Examples of double--double radiogalaxies from the study of~\cite{Mahatma19}. The~yellow contours show the emission at 150 MHz from LOFAR while the grey scale represent the high frequency, high spatial resolution emission from follow-up VLA observations. The~combination of the two highlights the presence of two sets of lobes.  See~\cite{Mahatma19} for~details. Reproduced with permission from Astronomy \& Astrophysics, \copyright ESO.}

\end{figure}

The prototypical restarted radio AGN are considered to be the DDRG. They are identified by two (or in some rare cases, three~\cite{Brocksopp07,Chavan23}) pairs of radio lobes and characterised by an extremely stable axis of the ejecta (see~\cite{Schoenmakers00,Konar13,Mahatma19}). The~large-scale (many hundred~kpc to Mpc) lobes are produced by past episode(s) of activity, while on the smaller scale are the more recently restarted lobes. 
The largest sample of DDRG is presented in~\cite{Mahatma19}. 
These sources were selected from LOFAR LoTSS 150 MHz images of the  HETDEX area \mbox{(400 sq deg)} and were followed up with higher-resolution VLA observations to confirm the structure of the inner lobes; see Figure~\ref{fig:DoubleDouble} for some~examples. 

However, as~mentioned in Section \ref{sec:sec2}, the~restarted activity can be identified also by other properties. 
Similar to what was performed for the remnant sources (and using the same reference sample of sources larger than 60 arcsec extracted from the LOFAR 150~MHz image of the LH region), the~study by~\cite{Jurlin20} selected, using multiple criteria, candidate restarted radio sources. Most of the restarted candidates  were selected based on their high CP combined with low surface brightness of the extended emission (as described \mbox{in Section \ref{sec:sec2}}). The~restarted nature of these objects was later confirmed by a follow-up study which made use of the LOFAR International Baselines image of the LH \cite{Sweijen22}, reaching a spatial resolution of 0.3 arcsec. This has allowed to characterise the central regions of these objects and to show~the presence of small jets seen in the central region while being surrounded by low surface brightness and amorphous emission on large scales; see~\cite{Jurlin23}  for details.
DDRG identified in the LH region by visual inspection were found to represent a minority of the group of candidate restarted sources.  
Taking all the candidates together, the~study of~\cite{Jurlin20,Jurlin23} derived an incidence of about 13\% of restarted radio~sources.

\begin{figure}[H]
\begin{adjustwidth}{-\extralength}{0cm}
\centering
\includegraphics[width=6.5cm,angle=-90]{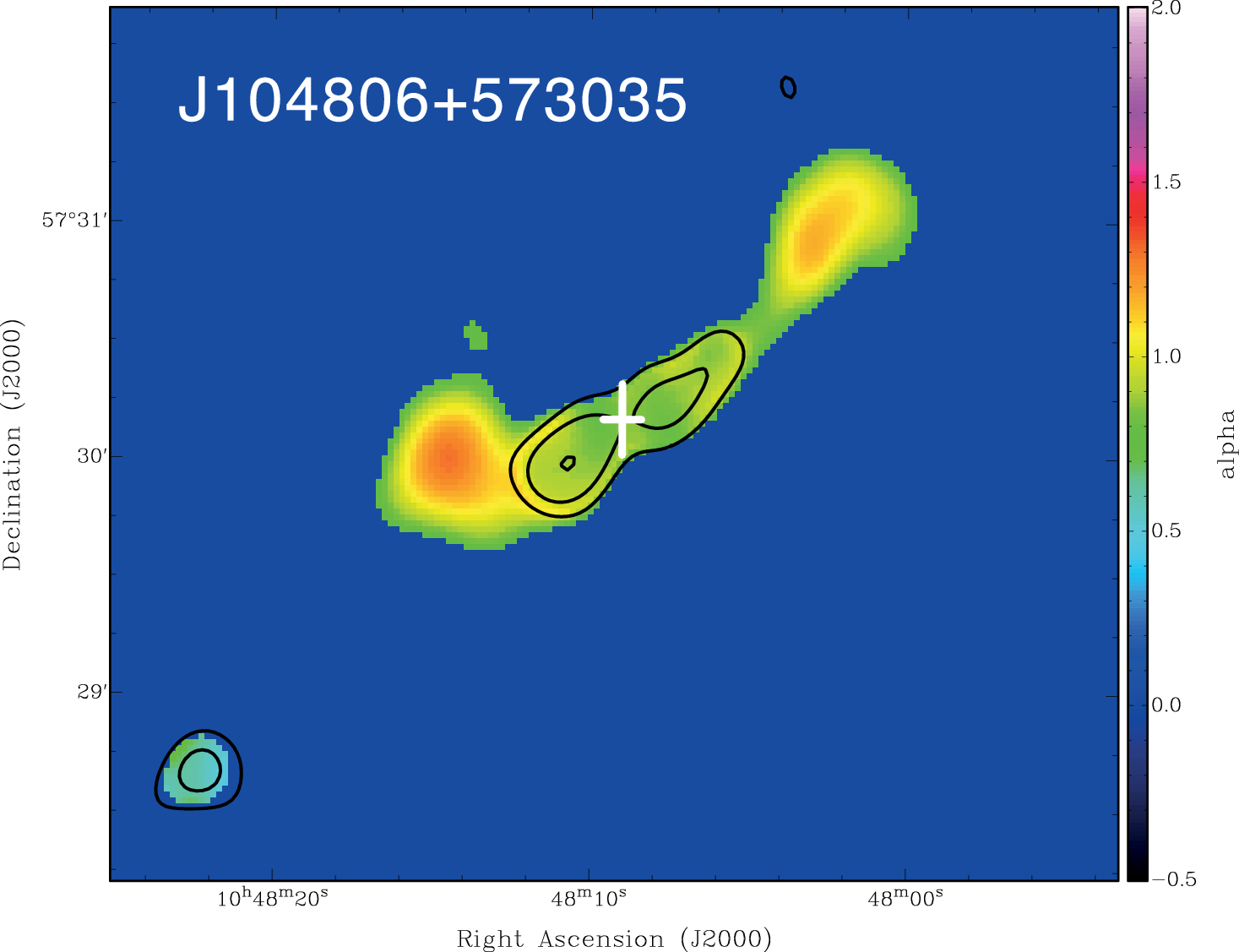}
\includegraphics[width=6.5cm,angle=-90]{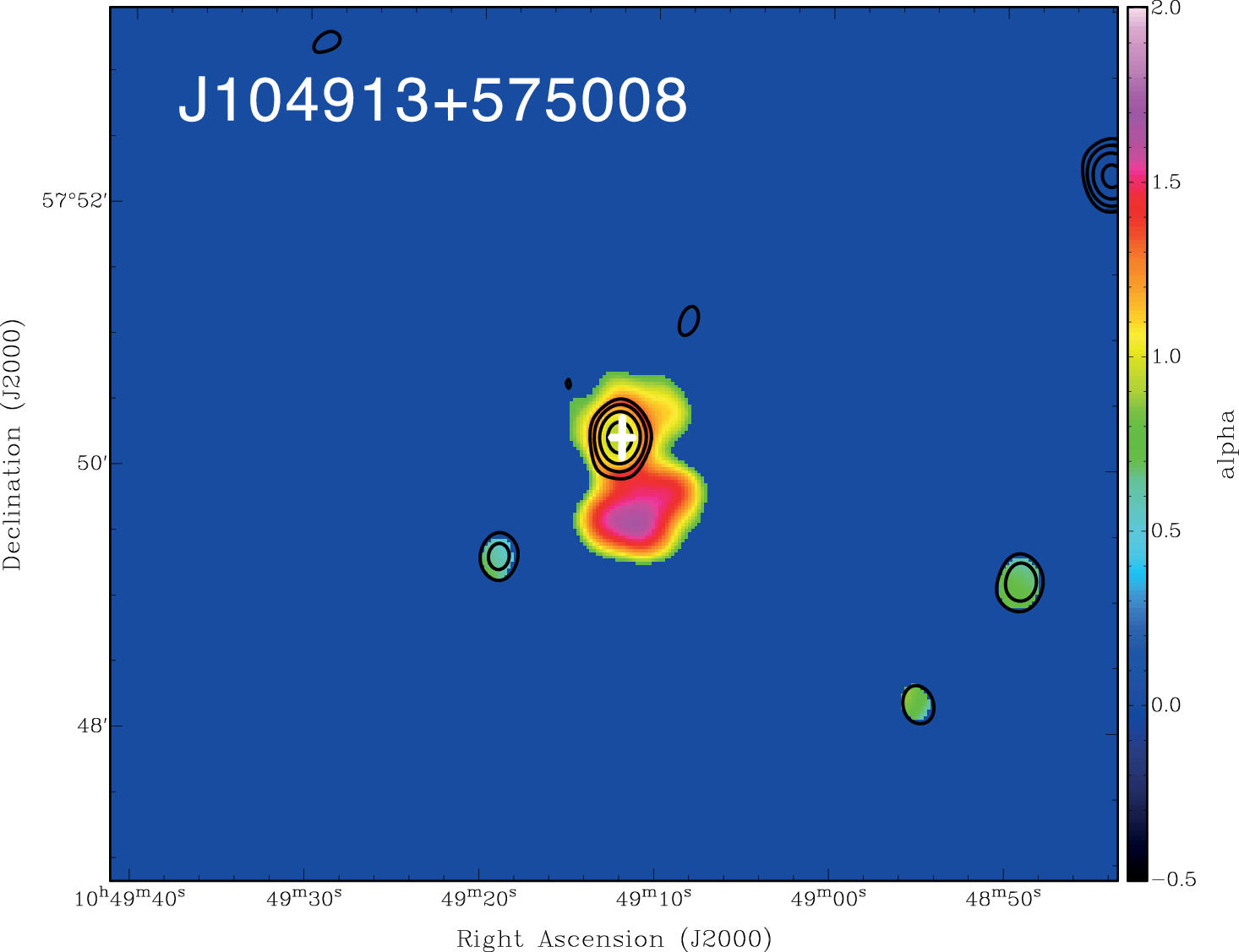}
\end{adjustwidth}
\caption{\label{fig:SI-restarted}Examples taken from~\cite{Morganti21b} of restarted radio galaxies identified thanks to the properties of their spectral index between 150 and 1400 MHz. The~red regions represent spectral indices steeper than 1.2 (i.e., USS), yellow regions represent a spectral index around 0.8. The~yellow edge around the source is an artefact due to the different sensitivity of the two observations (LOFAR and Apertif; see~\cite{Morganti21b} for details). Reproduced with permission from Astronomy \& Astrophysics, \copyright ESO.}

\end{figure}

Useful for identifying restarted radio sources are also the resolved spectral index images. These were produced in the LH region by combining 150~MHz LOFAR and 1400~MHz Apertif images (see~\cite{Morganti21a,Morganti21b}), the~latter providing higher spatial resolution \mbox{($12^{\prime\prime} \times 12^{\prime\prime}/ sin \delta$)} and higher sensitivity (noise of about {40} 
 \upmuJybeam) than other surveys at the same frequency like the NVSS.
A number of objects were found having a central active region (with $\alpha_{1400}^{150} \sim 0.7$) surrounded by a diffuse, low-surface brightness characterised by USS emission. Two examples are shown in Figure~\ref{fig:SI-restarted}. 
In these objects, the old (remnant) phase of activity is still visible while a new phase has already re-started. Therefore, they are particularly useful for deriving the time scales of the life cycle of radio sources. 
Until recently, only one example of restarted radio source identified by the spectral properties was known \mbox{(i.e., 3C~388,~\cite{Roettiger94,Brienza20})}, while more cases have now been found by building spectral index images of large areas (see~\cite{Morganti21a,Morganti21b,Kutkin23} for a more recent example).

While the above samples of remnant and candidate restarted sources include only large sources (in order to resolve their morphology using the available surveys), observations of the young and small-size radio sources are also relevant for understanding the variety of time scales associated with the dying and restarting phases. These studies have shown the existence of not only a population of young dying sources (with \mbox{ages $< 10^6$ yrs,~\cite{Kunert06,Webster21,Orienti23})} but~also cases of restarted sources in young remnant objects \cite{Orienti23}{.} These findings are consistent with the presence, at~least in some sources, of~a rapid duty cycle, i.e.,~restarting activity before the emission of the remnant structure fades~completely. 

Thus, the~studies so far show that a variety of time scales characterise the restarting activity. Indeed, the~time scales can be very short, resulting, in~the extreme cases, in~a ``flickering'' of the radio emission. An~example of this flickering (on Myr time) emission is the case of Fornax A~\cite{Maccagni20}. These extreme situations are difficult to trace by the available radio surveys, and instead detailed multi-wavelength observations are~needed. 

Finally, it is worth mentioning that the derived ages of the restarted jet can be affected by the presence of interaction between the radio plasma jet and the surrounding medium. 
In these cases, the~flow in the jet can be temporarily disrupted by~the rich medium in which they expand, and~the derived ages should be taken with care (see Ref.~\cite{Kukreti22} and the references therein). On~the other hand, these cases are relevant for quantifying the impact that a new phase of radio AGN can have on the surrounding gas (see Section \ref{sec:LifeCycle}).


\section{Implications for the Timing of the~Life Cycle}
\label{sec:Intepretation}

\begin{figure}[H]
\begin{adjustwidth}{-\extralength}{0cm}
\centering
\includegraphics[width=15.5cm]{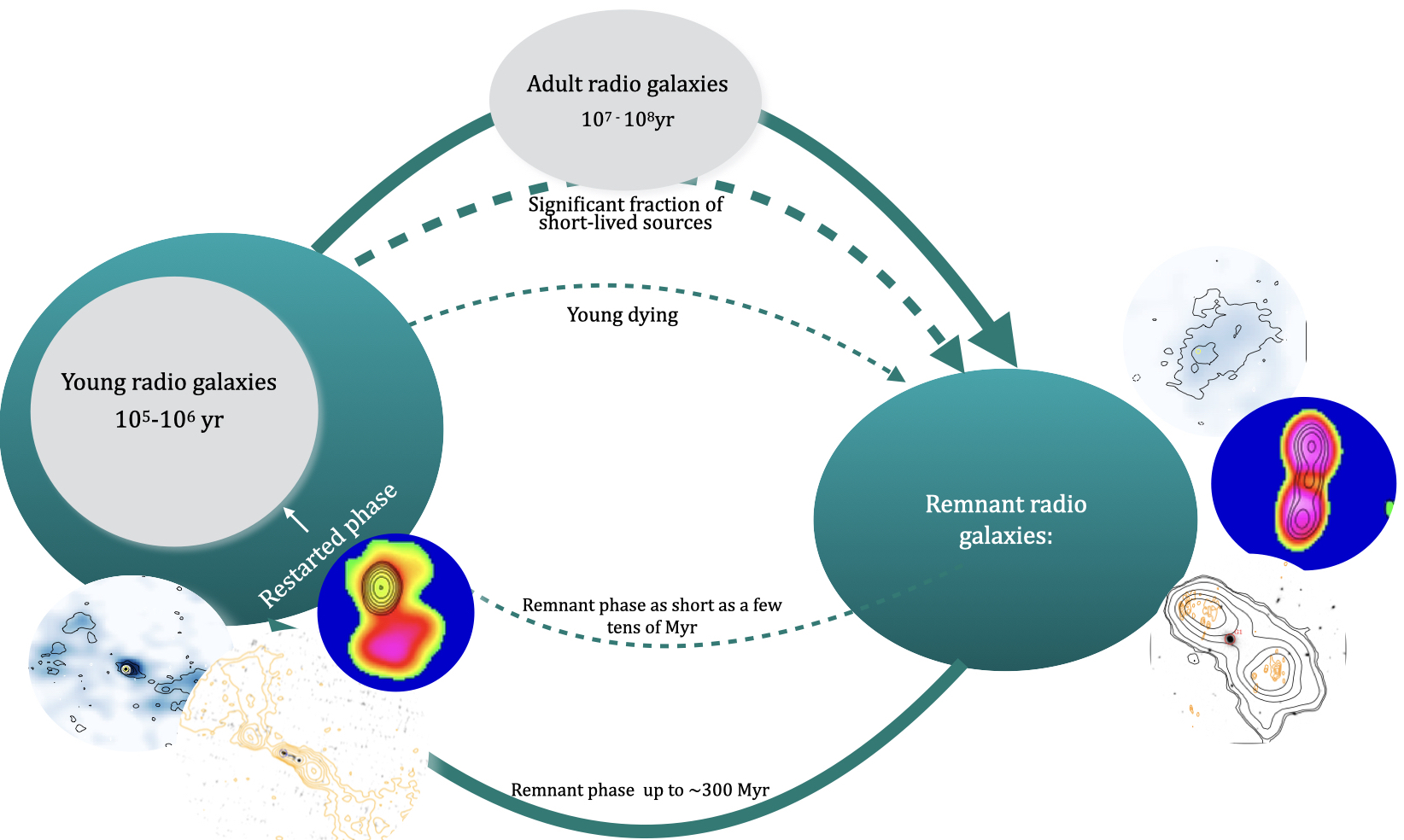}
\end{adjustwidth}
\caption{\label{fig:CartoonCycle} {Illustration} 
 {{summarising} the main phases of the life cycle of a radio source with some of the time scales added based on the studies discussed in this paper and modified from~\cite{Morganti21a}. The~existence of ``young dying'' has been discussed in~\cite{Kunert06,Orienti23} and the references therein. The~significant fraction of short-lived sources is the result of the simulations of~\cite{Shabala20}, see Section~\ref{sec:Intepretation} for details.} Examples of the objects (remnant and restarted radio sources) used to derive the time scales {({some} of them discussed in this paper)}  are also~shown.}

\end{figure}

The results described above, schematically summarised in the illustration in Figure~\ref{fig:CartoonCycle}, have already expanded our view of the life cycle of radio~sources.

\begin{figure}[H]
\hspace{-57pt}\includegraphics[width=0.9\linewidth]{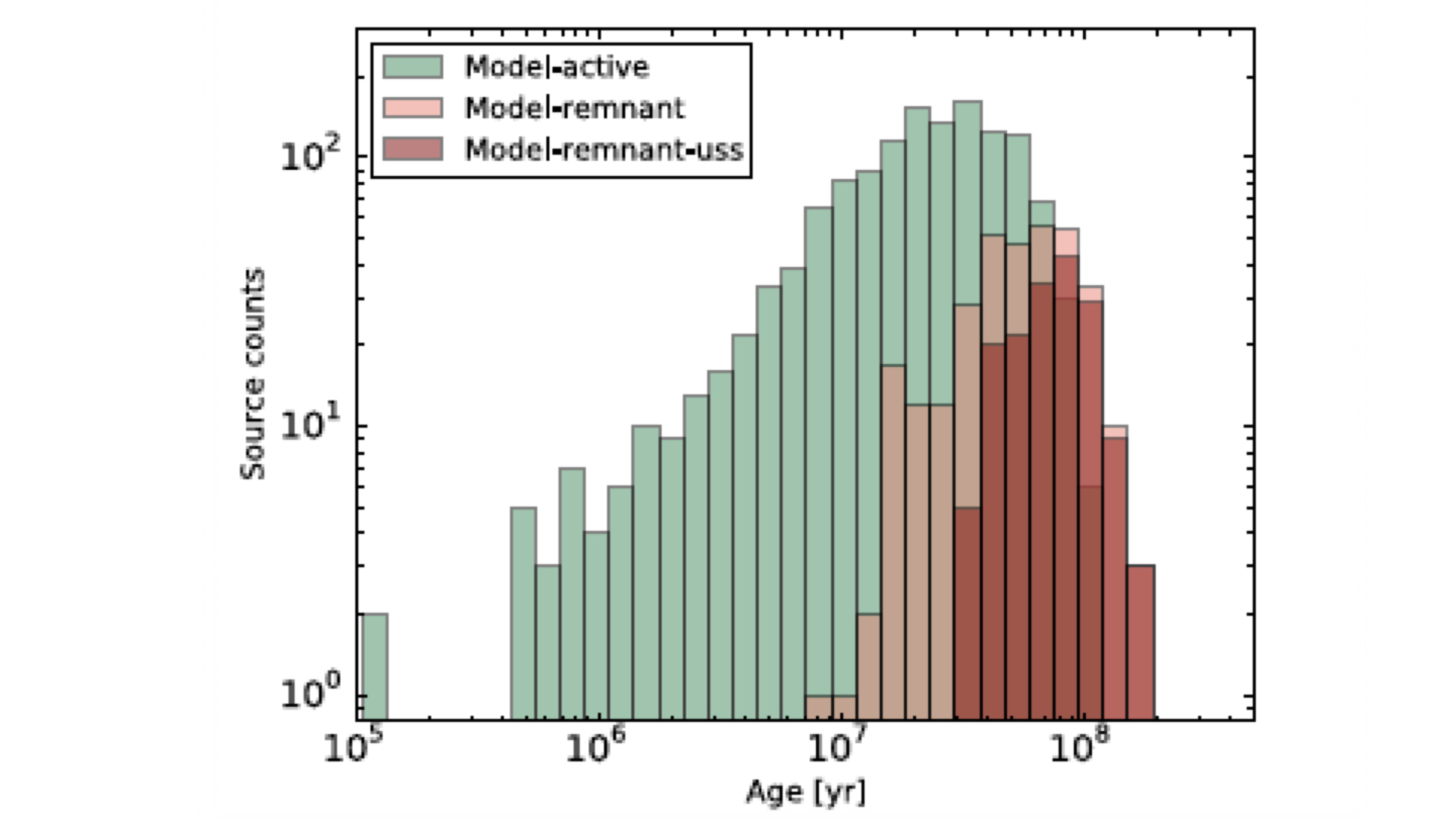}
\caption{\label{fig:MockCatalogue} {Distribution} 
 of ages which better describes the observations obtained from a mock catalogue, see details in~\cite{Brienza17}. Green marks active sources. Remnant sources (selected either from their morphology or their radio spectrum)  are indicated in brown. Most of the remnants are “young” (light brown), i.e.,~in the phase shortly after the switching off (i.e., a few $\times 10^7$ yr) and their low-frequency spectrum is not yet ultra steep. Remnants selected based on their steep, low-frequency spectrum are the older remnants in the sky (ages $> 10^8$ yrs). }
\end{figure}

The fraction of remnants sources is  consistently found in~radio selected samples to be $\lesssim$10\%.  
This relatively small fraction suggests that the remnant emission evolves quickly, and that fades rapidly once the core/jets turn off \cite{Mahatma18}.
However, other findings suggest that the situation is likely more complicated than this. Somewhat surprisingly, the~number of USS has not increased by observing at low frequencies: only about 4\% of the remnants are USS. Instead, most of the studies indicate that remnants radio sources are characterised (at low frequencies) by a range of spectral indices. Although~it may look surprising, this is consistent with the presence of remnants with a distribution of ages. 
This has been confirmed by the comparison of the results of the observations of the LH region (described above) with mock catalogues constructed to reproduce---in term of sensitivity---these observations (see~\cite{Godfrey17,Brienza17}). Interestingly, the~number of remnant sources observed can be reproduced by the mock catalogues only if in the dying phase  adiabatic expansion losses are included in addition to  radiative losses. This suggests that the radio lobes are still overpressured (with respect to the ambient medium) when reaching the off phase and implies that the expansion of the lobes continue during the remnant phase with a consequent (fast) evolution and dimming of the remnant emission (see the modelling presented in~\cite{Brienza17,Godfrey17}). 
The distribution of the ages of the sources in the mock catalogue is shown in Figure~\ref{fig:MockCatalogue} (see~\cite{Brienza17} for details). {{Table~5 in} 
\cite{Brienza17} shows the fraction of remnant radio sources in the mock catalogue: the models, including radiative and expansion losses predict that all remnant sources can be selected by the USS criteria only if the spectra can be built covering from 150 to, for example, 5000 MHz. For~spectra extending only to 1400 MHz, half of the remnant sources are not USS; therefore, other criteria (like the morphology) are needed to identify them}. Indeed, the~histogram shows that some remnants have just switched off (light brown in the figure), while others are old remnants (dark brown). Most of the remnants are “young”, i.e.,~in the phase shortly after the switching off \mbox{(i.e., a few $\times 10^7$ yr)}. This means that the low-frequency spectrum is not yet ultra steep. Remnants selected based on their ultra-steep, low-frequency spectrum represent the older remnants in the sky (ages $> 10^8$ yrs).
Indeed, the~USS emission observed  at low frequencies (that in some of the remnant sources can extend up to few $\times 100$ kpc) suggests that the older remnant sources have ages up to 160--300  Myr, if~a frequency break between \mbox{600 and 150~MHz} and the equipartition magnetic field  B${\rm eq} = 3$ $\upmu$G are assumed~\cite{Morganti21a}. {{Because} they are dependent on the magnetic field (notoriously difficult to derive) and from the expansion of the plasma, which is also difficult to quantify, these numbers should be taken with care (see also discussion in}~\cite{Hardcastle20}).
\vspace{-4pt}

It is also interesting to consider the presence of faint cores in some of the {{bona fide} 
} remnant structures (i.e., in sources with USS extended emission) when deep enough high-resolution, high-frequency observations are obtained.
As mentioned above, two hypotheses have been proposed to explain this: either some cores never really switch off (but only dim in flux), or we are seeing the beginning of a new phase of activity. 
The latter scenario has interesting implications for the timing of the restarted phase. 
All these properties can be explained if the ``off'' time of the radio AGN has not been longer than a few tens of Myr, i.e.,~the remnant emission had not {{enough}} time to fade before the new phase of activity started. 
It is worth noting that a relatively fast cycle of activity has been also suggested for the DDRG. The~studies by~\cite{Konar13,Orru15} have shown that the ``off'' phase in these sources tends to be comparable or shorter than their on phase, ranging from a few Myr to a few tens of Myr.  {{However,} as~a word of caution,  we need to keep in mind that observations may favour the detection of DDRG with short off times, where the old lobes have not faded yet}.
\vspace{-6pt}


The new results from the observations of remnant and restarted radio AGN have stimulated new developments of the theoretical modelling of the evolution of radio sources (see~\cite{Hardcastle18,Hardcastle20,Shabala20,Turner23}). 
For example,~\cite{Shabala20} have used RAiSE (Radio AGN in Semi-Analytic Environments,~\cite{Turner15}) to compare the predictions of the simulations with the results from the observations obtained in the LH region and described above \cite{Brienza17,Jurlin21}{.} The~strategy used has been to 
(i) use a sample of active radio galaxies \cite{Jurlin21}{ }with host information to derive their physical parameters using RAiSE (i.e., sampling the distribution of physical properties of the radio galaxy population); (ii) employ outputs of galaxy formation models to quantify jet environments; and (iii) use these distributions to make predictions for remnants. 
The main result obtained from the comparison with the LH study is that the age distribution of radio galaxies most likely follows a power law: short-lived jets should exist to obtain enough remnant lobes which are still visible \cite{Shabala20}{.} 
As mentioned above, other observational findings seem to confirm this result \cite{Orienti23}.  

While these simulations describe the general occurrence of the remnant and restarted phases, numerical simulations are also in the process of being developed for describing in more detail the morphology and dynamics of the observed radio sources. 
Particularly challenging is the modelling of the restarted phase. The ``off'' time is a key parameter for the simulations in order to reproduce the conditions of the ISM in which the restarted jet will expand (e.g., \cite{Safouris08,Walg14}). Thus, the~improved estimates of the length of the ``on'' and ``off'' phases derived from the new observations are providing key constraints to define the parameter space that needs to be explored by the new generation of~simulations.


\section{Life Cycle and~Gas}
\label{sec:LifeCycle}

Characterising the cycle of activity of radio AGN has broader implications, in~particular, for the understanding of the feeding and feedback of AGN. 
How this cycle is regulated, i.e.,~the process of triggering and stopping the  activity of the SMBH, and~how the starting (or restarting) of a jet impacts the surrounding ISM/IGM in the host galaxy, i.e.,~feedback effect, are two key open~questions. 

Although various processes have been suggested to explain the triggering of AGN (\mbox{see e.g.,~\cite{Ramos12,Storchi19,Combes23} {{for reviews}})}, what actually originates the stop-and-start of the radio emission (and on which time scales) it is not fully understood.   To~explain the short-lived young radio sources,~\cite{Czerny09}  suggested a radiation pressure instability scenario which may cause an intermittent activity of the central engine: outbursts of radio emission, with~a duration of \mbox{$10^{3-4}$ years}, repeated regularly every $10^{5-6}$ years.  To~explain the longer cycle of activity,  the~availability of gas, perhaps connected to chaotic cold accretion~\cite{Gaspari17}, and~the self-regulating feedback cycle, could be possible processes.  Interestingly, some studies have suggested that young or restarted radio sources are more rich in gas, \mbox{e.g.,~\cite{Holt08,ODea21},} while powerful radio galaxies seem to be more likely hosted by merger or \mbox{interacting galaxies~\cite{Ramos12}.}

\begin{figure}[H]
\centering
\includegraphics[height=0.4\linewidth]{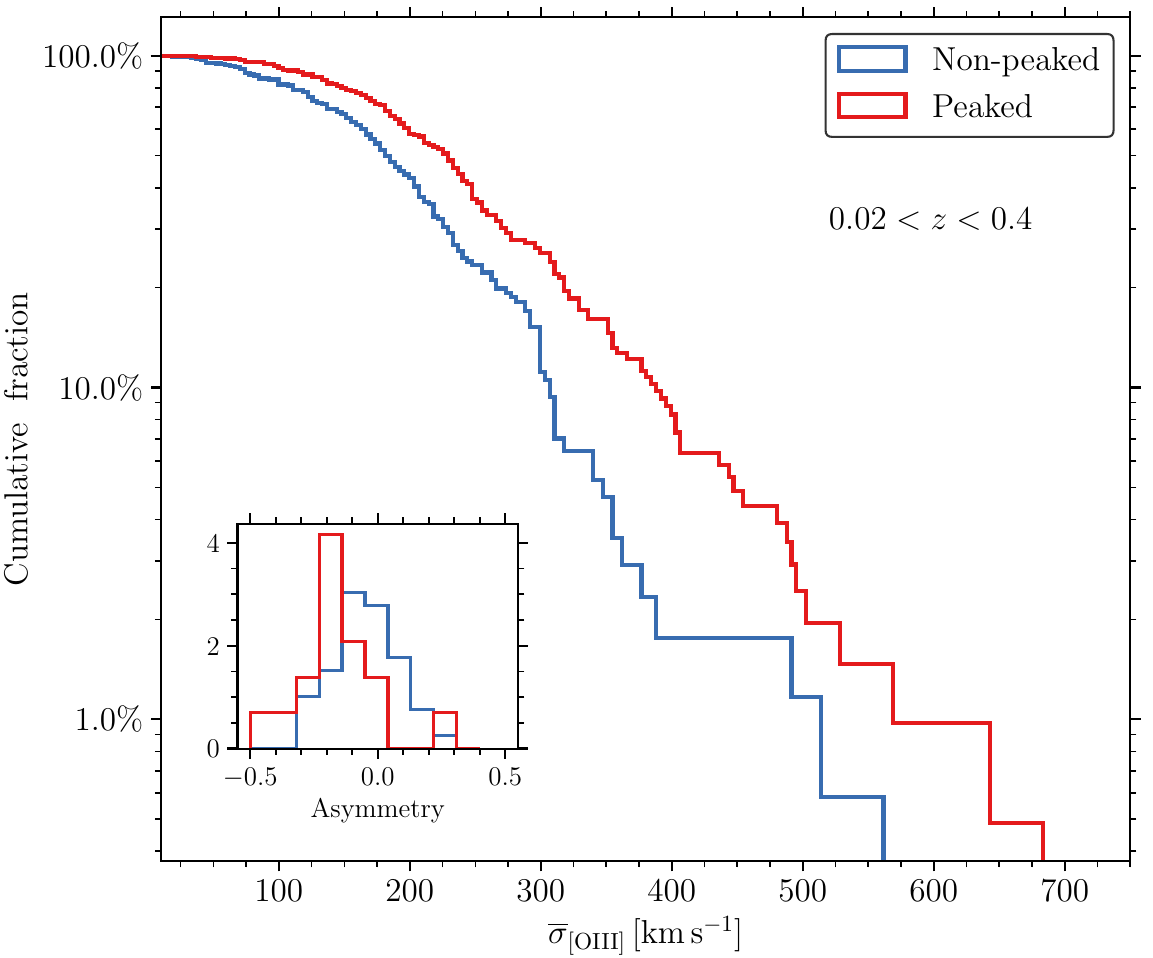}
\includegraphics[height=0.41\linewidth]{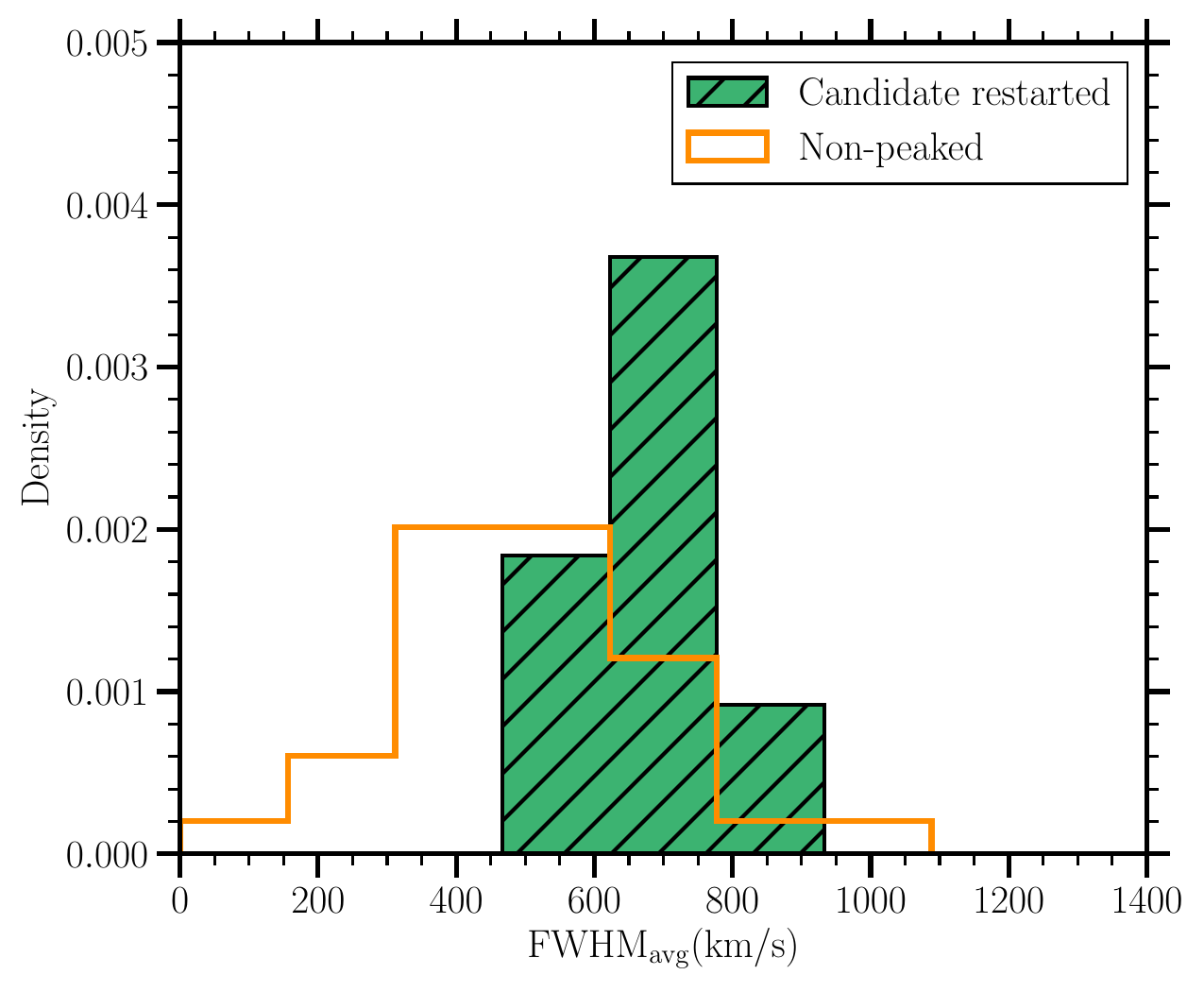}
\caption{\label{fig:Peaked-OIII} Left: Cumulative distribution of the [OIII] lines width (obtained from SDSS spectra) for peaked and non-peaked sources from the sample of {Kukreti~et~al.,} in prep. Right: Average FWHM distributions of the [OIII] profiles of candidate restarted and the non-peaked radio AGN. Plots taken from~\cite{Kukreti23}; see this paper for~details. }
\end{figure}

More progress have been made in the investigation of the impact of 
radio jets on the surrounding medium. While the main role of radio jets for feedback is usually considered to be preventing the cooling of the gas from the cluster scale~\cite{McNamara12}, evidence has emerged that radio jets, like radiation and winds, can also originate gas outflows on galaxy scale. Simulations indicate that this can be more efficiently performed by a jet in its initial phase of evolution \cite{Sutherland07,Mukherjee18} (or in the restarted phase).  Single object studies have also confirmed that this indeed can happen (see {\cite{Holt08,Shih13,Molyneux19,ODea21}}). A~statistical study specifically aimed at selecting young radio sources based on the peaked radio spectrum has further confirmed these results (see~\cite{Kukreti23} and Figure~\ref{fig:Peaked-OIII}).   These results suggest that the young and restarted phases are those where the gas in the nuclear region of the host galaxies is mostly affected by the radio jet. 

These results do not exclude the effect of radiation/wind but show that when a young (or restarted) jet is present, the impact on the gas is visible. This impact may last for a short time (order of $10^6$ yr) as seems to emerge from the actual size of the outflows (typically limited to the kpc scale).
\vspace{-6pt}
\unskip
\section{Conclusions and Future~Possibilities}
\label{sec:Conclusions}

This paper has briefly summarised some of the recent progresses made, thanks to the availability of new surveys, in~the study of the life cycle of radio AGN. The~view of the two most elusive phases, remnant (dying) and restarted, has significantly improved: these phases appear now richer and more complex than previously thought.
Low radio-frequency surveys (i.e., using LOFAR, MWA and uGMRT) together with the expanded availability of ancillary data (i.e., optical identifications and redshifts) have opened up the possibility of selecting larger samples of sources in these elusive phases. Crucially, this can now be performed using multiple selection criteria: the results have demonstrated that this is essential for a more comprehensive census of remnant and restarted radio sources. 
The studies described in this paper provide the starting point and now can be expanded to larger areas,~pushed to fainter targets. This will provide better statistics, covering a larger parameter space in terms of radio luminosity and stellar mass of the host galaxy, and~better constraints for the models of radio galaxy~evolution.

However, performing these larger selections provides a challenge in itself.
With many million sources to classify, the~development of advanced software tools for an automatic selection and classification of the objects is needed. This is a rapidly expanding field \mbox{(e.g.,~\cite{Galvin20,Ndung'u23})} which will also benefit from tools specifically tuned to identify rare objects like remnant radio source (see~\cite{Mostert23} and {Brienza~et~al.,
} 
in prep). 

Many other new opportunities for improving our knowledge of the life cycle are already available (or will be soon). For~example, these studies can be further expanded by using spectral index images, which can complement the analysis of the morphology and CP described above.  As~shown in this paper, surveys at GHz frequencies can be used together with the low-frequency ones. The~results so far have also shown the importance of {{being able}} to achieve a good spatial resolution in order to separate the central and the extended emission.  
At low frequencies, LOFAR is also providing new possibilities for these studies. The~LOFAR LBA Sky Survey (LoLSS) at 50~MHz (see~\cite{deGasperin23}) will become available soon at full resolution. This will allow to build spectral index images extended to this very low frequency. A~proof of concept has been performed in the Bo\"otes region  with spectral indices obtained at three frequencies (50, 150 and 1400 MHz) at about 15 arcsec resolution \cite{Kutkin23}{.}  
Other exciting opportunities are given by the high resolution that can be achieved by the LOFAR International Baselines Array. Thanks to the improved pipeline~\cite{Morabito22,Sweijen22}, images at 150 MHz with $\sim$0.3 arcsec resolution will become routinely available and will help identify and characterise the nuclear regions of radio AGN as shown for the LH area by~\cite{Jurlin23}.

The expanding possibilities for obtaining optical identifications for the radio sources and characterise the properties of the host galaxy are a major, key addition. The~optical identification of the entire LoTSS DR2 has been recently published \cite{Hardcastle23}{ } and provides a great database of crucial ancillary information. These data  also improve our knowledge of the environment, which is needed for a complete understanding of the parameters impacting the evolution of the radio sources. For~tracing the presence and conditions of the gas, Integral Field Unit (IFU) instruments which will perform large surveys are also becoming available (e.g.,  WEAVE~\cite{Jin23} and 4MOST~\cite{Jong22}) and will play an important~role. 

In summary, a~wealth of new data are opening up great possibilities for improving our understanding of the life cycle of radio AGN. Handling the data will provide major challenges as well as making sure theoretical models for interpreting the data are becoming available. Already from the results presented in this paper, it is clear that this effort will be rewarding in terms of scientific results, and it will also make sure the community is ready for the next big step ahead that will be made possible by the Square Kilometre Array (SKA).

\vspace{6pt} 


\funding{This research received no external funding. 
}

\dataavailability{\textls[10]{Data from the surveys discussed in this paper (LOFAR, Apertif, MWA,~etc.) are available mostly through the Virtual Observatory.} }

\acknowledgments{I would like to thank Marisa Brienza, Nika Jurlin, Stas Shabala and the LOFAR and Apertif Survey Teams for the help, comments and discussions. Some of the results presented in this overview have only been possible thanks to their~contribution.}

\conflictsofinterest{The author declares no conflicts of~interest.} 


\abbreviations{Abbreviations}{
The following abbreviations are used in this manuscript:\\
\vspace{-9pt}
\begin{longtable}[l]{@{}lll}
AGN& Active galactic nucleus\\
Apertif & APERture Tile In Focus \\
ASKAP & Australian Square Kilometre Array Pathfinder \\
ATCA & Australian Telescope Compact Array \\
CP& core prominence, S$_{core}$/S$_{total}$\\
DDRG & Double--double radio galaxy \\
FIRST&Faint Images of the Radio Sky at Twenty centimetres\\
GLEAM & Galactic and Extra-Galactic All-Sky MWA Survey \\
uGMRT & upgraded Giant Metrewave Radio Telescope \\
HBA & High-band Array\\
LBA & Low-Band Array\\
LH & Lockman Hole region\\
LoLSS & LOFAR LBA Sky Survey\\
LOFAR & LOw Frequency ARray\\
LoTSS& LOFAR Two-metre Sky Survey\\
MWA & Murchison Widefield Array \\
NVSS& The NRAO VLA Sky Survey\\
SMBH & Super Massive Black Hole\\
TGSS & TIFR-GMRT Sky Survey \\
USS&Ultra Steep Spectra\\
VLA& Karl G. Jansky Very Large Array \\
VLASS & Very Large Array Sky Survey  \\
WSRT & Westerbork Synthesis Radio Telescope\\
\end{longtable}
}

\begin{adjustwidth}{-\extralength}{0cm}
\setenotez{list-name=Note}
\printendnotes[custom] 

\reftitle{References}


\PublishersNote{}
\end{adjustwidth}
\end{document}